\documentclass[
    ,final            
    ,numberedheadings 
  ]
  {aipproc}

\layoutstyle{6x9}
\usepackage{amsfonts}
\usepackage{bbm}
\usepackage{bm}

\newcommand{\beq}{\begin{equation}}
\newcommand{\eeq}{\end{equation}}
\newcommand{\ba}{\begin{eqnarray}}
\newcommand{\ea}{\end{eqnarray}}

\begin{document}

\title{Correlation Functions of Winding Strings\\ in
AdS$_{3}$ \footnote{Based on work in
collaboration with E. Herscovich and C. N\'u\~nez \cite{hmn,4p}.}}


\author{Pablo Minces}
{address={Instituto de Astronom\'{\i}a y F\'{\i}sica del Espacio
(CONICET - UBA), \\
C.C.67 - Suc. 28, 1428 Buenos Aires, Argentina.}}

\begin{abstract}
We review certain results for amplitudes of spectral flowed operators in
string theory on AdS$_{3}$. We present the modified Knizhnik-Zamolodchikov
and null vector equations to be satisfied by correlators including $w=1$
operators. We then discuss the three point function of two $w=1$ and one
$w=0$ operators in the $x$-basis, and perform a consistency check on
the definition of the $w=1$ operator. We finally exhibit the steps in the
calculation of the winding conserving
four point functions for operators in arbitrary spectral flowed sectors,
both in the $m$- and $x$-basis, under the only assumption that at least
one of the operators is in the spectral flowed image of the highest weight
discrete representation.
\end{abstract}

\maketitle

\section{Introduction}
In this article we consider certain correlation functions in the $SL(2,R)$
WZW
model, which describes strings propagating in the three
dimensional Anti-de Sitter space (AdS$_{3}$). The motivations are that to
study
such a model would allow to investigate aspects of the AdS/CFT
correspondence beyond the supergravity approximation, and that the
propagation of strings in this non-compact background is not
well-understood yet. The applications of this model to black
hole physics come as an additional motivation. However,
diverse difficulties arise due to the non-rational structure of this CFT.

Here we employ two different basis to calculate the amplitudes of the
theory. In the $x$-basis the isospin parameter is understood as the
coordinate of the boundary of the AdS$_{3}$ space, whereas in the
$m$-basis
the generators of the global isometry are diagonalized. The
transformation relating an operator $\Phi_{j;m,\bar
m}(z, \bar z)$ in the $m$-basis to a corresponding operator
$\Phi_{j}(x,\bar x)$ in the $x$-basis is given by

\beq
\Phi_{j; m, \bar m}(z,{\bar z}) = \int \frac{d^2 x}{|x|^2}\; x^{j-m}\,
\bar x^{j - \bar m}\, \Phi_{j}(x,\bar x;z,{\bar z}) \; .
\label{max}
\eeq

The states in the long strings sectors correspond to continuous
representations of $SL(2,R)$ with spin $j=\frac 12 + is$, $s\in$ {\bf
R}, whereas short string
sectors are discrete
representations with spin $j\in$ {\bf R} and unitarity bound
$\frac 12 < j < \frac{k-1}2$. The conformal weight of the primary fields
is given by $\Delta_j = -\frac{j(j-1)}{k-2}$, where $k>2$.

There is however an additional feature of the $SL(2,R)$ CFT which is the
fact that it contains different sectors which are labeled by the winding
number or spectral flow parameter $w\in{\bf Z}$.\footnote{We point out
that in the $x$-basis the
operators are labeled with positive $w$.} Thus we will also consider here
spectral flowed operators which we write $\Phi^{w,j}_{J,M;{\bar
J},{\bar M}}(z)$ in the $m$-basis, or $\Phi^{w,j}_{J,{\bar
J}}(x,z)$ in the $x$-basis. The conformal weight and spin are given by

\beq
\Delta^w_{j}=-\frac{ j( j-1)}{k-2}- m w
-\frac{k}{4}\;w^{2}\; ,
\label{520}
\eeq
with a similar expression for $\bar\Delta_j^{w}$, and

\beq
J=\;\mid M\mid\; =\;\mid m+\frac {kw}{2}\mid\,,\qquad \qquad\bar J =
\;\mid \bar M \mid\;
= \;\mid\bar
m + \frac
{kw}{2}\mid\; .
\label{1aa}
\eeq
In particular, a $w=1$ operator in the $x$-basis is obtained from a $w=0$
operator through \cite{malda3}
\ba
&&
\!\!\!\!\!\!\!\!\!\!\!\!\!\!\!
\Phi^{w =1,j}_{J,\bar J}\!(x,\bar x; z, \bar z) \equiv\!\!
\lim_{\epsilon,\bar \epsilon\rightarrow 0}\!
\epsilon^{m}\bar\epsilon^{{\bar m}}\!\!
\int d^{2}\!y\;
y^{j- m -1} \bar y^{j-{\bar m}-1} 
\Phi_{\!j}\!(x\!+\!y,\bar x \!+\!
\bar y \,;
z\!+\!\epsilon,
\bar z\! +\! \bar\epsilon)
\Phi_{\!\frac k2}\!\!(x,\bar x ; z, \bar z)\; ,
\nonumber\\
&&\!\!\!\!\!\!\!\!\!\!\!\!\!\!\label{1}
\ea
where $\Phi_{\frac k2}$ is the so-called spectral flow operator.

An important additional feature is that the amplitudes in WZW models
should satisfy the Knizhnik-Zamolodchikov (KZ) equations \cite{zamo2}
which for some $N$ point function $A_{N}$ in the $x$-basis read
\ba
&&\!\!\!\!\!\!\!\!\!\!\!\!\!\!
(k\!-\!2)\frac{\partial A_{N}}{\partial z_{i}}
=\!\!\!\sum_{n=1,\; n\not=
i}^{N}\frac{1}{z_{i}-z_{n}}{\Bigg [}\!(x_{n}\!-\!x_{i})^{2}
\frac{\partial^{2}}{\partial x_{i}\partial x_{n}}
\!+
2(x_{n}\!-\!x_{i})\!
\left(\!j_{n}\frac{\partial}{\partial
x_{i}}\!-\!j_{i}\frac{\partial}{\partial
x_{n}}\!\right)\!-\!2j_{i}j_{n}\!
{\Bigg ]}A_{N}\, ,\nonumber\\
&&\!\!\!\!\!\!\!\!\!\!\!\!\!\!
\label{15b}
\ea
(for $1\leq i\leq N$). In addition, we also point out that, since the
spectral flow
operator has a null descendant, any
correlation function including such an operator at, say, $(x_{l},z_{l})$
(with $1\leq l\leq N$) should obey the null vector equation of the form

\beq
0=\sum_{n=1,\; n\not=
l}^{N}\frac{x_{n}-x_{l}}{z_{l}-z_{n}}\left[(x_{n}-x_{l})
\frac{\partial}{\partial x_{n}}+2j_{n}\right]\; A_{N}\; .
\label{16b}
\eeq

In this work we will comment on the generalization of the KZ and
null vector equations to the case of amplitudes involving $w=1$
operators in the $x$-basis, and on certain three and four point
functions including spectral flowed operators, which were computed in
\cite{hmn,4p}. The article is organized as follows. In section 2 we
consider the KZ and null vector equations as explained above. In section 3
we focus on three point functions
involving $w=1$ operators whereas in section 4 we consider four point
functions in a somehow more general way, since we deal with operators in
arbitrary spectral flow sectors and in both the $x$- and
$m$-basis.\footnote{See \cite{malda3,nunez1} for the computation of
the two point function for operators in arbitrary
spectral flow sectors.} Throughout this work we follow the conventions
and notation in \cite{malda3}.

\section{KZ and Null Vector Equations}

In order to obtain the generalizations of (\ref{15b}) and (\ref{16b}) to the
case of an $N$ point function including one $w=1$ operator and $N-1$
unflowed
operators, the proposal in \cite{hmn} was to start from an amplitude of
$N+1$ unflowed operators, one of which was required to be a spectral flow
operator. This should satisfy (\ref{15b}) with the replacement of $N$
by $N+1$. Then, by performing the operation (\ref{1}) in order to
construct ourselves a $w=1$ operator inside the correlator, we should be
able to find the generalization of (\ref{15b}) mentioned above. The result
in \cite{hmn} was
\ba \label{al2}
(k\!-\!2)\frac{\partial A^{w}_{N}(J)}{\partial z_{i}}\!\!\!&=&\!\!\!
-\left(j_{1}\!-\!J\!+\!\frac{k}{2}\!-\!1\right)\frac{x_{2}-x_{i}}{(z_{i}-z_{2})^{2}}\!
\left[(x_{2}-x_{i})\frac{\partial}{\partial x_{i}}-2j_{i}\right]
\!A^{w}_{N}(J+1)
\nonumber\\[1em]
&&
\!\!\!\!\!\!\!\!\!\!\!\!\!\!\!\!\!\!\!\!\!\!\!\!\!\!\!\!
\!\!\!\!\!\!\!\!\!\!\!\!\!\!\!\!\!\!\!\!\!\!\!\!\!\!\!\!
+\,\,
\frac{1}{z_{i}-z_{2}}\!{\Bigg [}\!(x_{2}\!-\!x_{i})^{2}\!
\frac{\partial^{2}}{\partial x_{i}\partial x_{2}} +
2(x_{2}\!-\!x_{i})\left(\!J\frac{\partial}{\partial
x_{i}}\!-\!j_{i}\!\frac{\partial}{\partial
x_{2}}\!\right)\!-\!2j_{i}J
{\Bigg ]} A^{w}_{N}(J) \\[1em]
&&
\!\!\!\!\!\!\!\!\!\!\!\!\!\!\!\!\!\!\!\!\!\!\!\!\!\!\!\!
\!\!\!\!\!\!\!\!\!\!\!\!\!\!\!\!\!\!\!\!\!\!\!\!\!\!\!\!
+ \!\!\!
\sum_{n=3,\; n\not=i}^{N+1}\frac{1}{z_{i}\!-\!z_{n}}{\Bigg [}\!(x_{n}\!-\!x_{i})^{2}
\frac{\partial^{2}}{\partial x_{i}\partial x_{n}}
+\; 2(x_{n}\!-\!x_{i})\left(\!\!j_{n}\frac{\partial}{\partial
x_{i}}\!-\!j_{i}\frac{\partial}{\partial x_{n}}\!\right)\!-\!2j_{i}j_{n}
{\Bigg ]}A^{w}_{N}(J) \nonumber
\ea
(for $3\leq i\leq N$). Here $J$ is the spin of the $w=1$ field. Notice
that the $w=1$ operator is possitioned at $(x_{2},z_{2})$ and the
possition $(x_{1},z_{1})$ is no longer present, since it had been
occupied by
the unflowed operator with spin $j_{1}$ which was then fussed with the
spectral flow operator at $(x_{2},z_{2}$). In the expression above,
$A^{w}_{N}(J)$ stands for
the amplitude of one $w=1$ operator and $N-1$ unflowed operators, and the
notation $A^{w}_{N}(J+1)$ means that we replace $J\longrightarrow J+1$ in
$A^{w}_{N}$, so that (\ref{al2}) is actually an iterative
expression in the
spin of the spectral flowed field. We point out that a similar
equation holds for the antiholomorphic part, where ${\bar J}$ turns out
to be the iterative variable.

Since $A^{w}_{N}$ was obtained out of an $N+1$ point function including
one spectral flow operator, then it must also satisfy a generalization of
(\ref{16b}). This is obtained through a similar procedure as above giving
\cite{hmn}
\ba
\left(j_{1}+J-\frac{k}{2}-1\right)\;A^{w}_{N}(J-1)=
\sum_{n=3}^{N+1}\frac{x_{n}-x_{2}}{z_{2}-z_{n}}\left[
(x_{n}-x_{2})\frac{\partial}{\partial x_{n}}+2j_{n}
\right]\;
A^{w}_{N}(J)\; ,
\label{al4}
\ea
which is iterative as well. As before, a similar expression exists for the
antiholomorphic part.

In this way, Eqs.(\ref{al2}) and (\ref{al4}), respectively generalize
(\ref{15b}) and (\ref{16b}) to the case of an $N$ point function including one
$w=1$ operator. The case of an $N$ point function including, say, $M$
$w=1$
operators and $N-M$ unflowed operators (with $1\leq M\leq N$) is obtained
in a similar manner by starting from an $N+M$ point function including $M$
spectral flow operators. This gives expressions which are iterative in the
spins of the $M$ $w=1$ operators in the resulting $N$ point functions.

It would be interesting to establish the connection between the results
above and those in \cite{ribault} regarding the KZ equation for amplitudes
of spectral flowed operators, where the calculations were performed
in a different basis than here.

\section{Three Point Functions}

Here we summarize the already known examples of three point
functions involving
$w=1$ operators, in the $x$-basis, before considering the four point
functions in the next section. The three point function of one $w=1$
and two $w=0$ operators was computed
in \cite{malda3}. It is given by
\ba \label{3w}
&&
\!\!\!\!\!\!\!\!\!
\left\langle\Phi^{w=1,j_{1}}_{J,{\bar J}}(x_{1},z_{1})\,
\Phi_{j_{2}}(x_{2},z_{2})\, \Phi_{j_{3}}(x_{3},z_{3})\right\rangle \sim\,
\frac{\Gamma\left(j_{1} + J - \frac{k}{2}\right)}{\Gamma\left(1 + J -
j_{2} - j_{3}\right)}\; \frac{\Gamma\left(j_{2} + j_{3} -
{\bar J}\right)}{\Gamma\left(1 - j_{1} - {\bar J} + \frac{k}{2}\right)}
\nonumber\\[.8em]
&&\qquad\quad \times ~~
\frac{1}{V_{conf}}\, B(j_{1})\, C\left(\frac{k}{2} - j_{1},
j_{2}, j_{3}\right)\, \pi\, \frac{1}{\gamma\left(j_{1} + j_{2} + j_{3}
- \frac k 2\right)} \nonumber\\[.8em]
&&\qquad\quad \times ~~
\left(x_{21}^{j_{3}-j_{2}-J}x_{31}^{j_{2}-j_{3}-J}
x_{32}^{J-j_{2}-j_{3}}\right)\left(
z_{21}^{\Delta_{3}-\Delta_{2}-\Delta^{w=1}_{1}}
z_{31}^{\Delta_{2}-\Delta_{3}-\Delta^{w=1}_{1}}
z_{32}^{\Delta^{w=1}_{1}-\Delta_{2}-\Delta_{3}}\right)
\nonumber\\[.8em]
&&\qquad\quad \times ~~ ({\rm antiholomorphic\; part}) \; ,
\ea
where $B(j)$ and $C(j_{1},j_{2},j_{3})$ are the coefficients
corresponding to the amplitudes of two and three unflowed operators
respectively. They were first obtained in \cite{tesch2} for the
$SL(2,C)/SU(2)$ model and then analytically extended to $SL(2,R)$ in
\cite{malda3} (see this reference for their expressions in our
conventions). Here $V_{conf}$ is the volume of the conformal group
of $S^{2}$ with two point fixed, and $\gamma (x)=\frac{\Gamma (x)}{\Gamma
(1-x)}$.

The following step in the literature was to compute the amplitude of two
$w=1$ and
one $w=0$ operators in the $x$-basis. Such a calculation was performed in
\cite{hmn}. The
starting point was the five point function of three unflowed operators
with two
spectral flow operators, which we call $A_{5}$. The corresponding
dependence on the $x_{i}$
coordinates was given in \cite{zamo4} (see also \cite{malda3}). Then, the
$z_{i}$ dependent part was computed in \cite{hmn} thus giving the
following complete expression
\ba
A_5\!\!&=&\!\!
B(j_1)\, B(j_3)\, C \left ( \frac k2 - j_1, j_2,
\frac k2 - j_3\right )
|z_{12}|^k |z_{13}|^{-2j_1}|z_{14}|^{-2j_2}|z_{15}|^{-2j_3} \nonumber\\[.8em]
&&\;\;\times ~~|z_{23}|^{-2j_1}|z_{24}|^{-2j_2}|z_{25}|^{-2j_3}
|z_{34}|^{2(\Delta_3-\Delta_1-\Delta_2)}
|z_{35}|^{2(\Delta_2-\Delta_1-\Delta_3)}
|z_{45}|^{2(\Delta_1-\Delta_2-\Delta_3)}\nonumber\\[.8em]
&&\;\; \times ~~|x_{12}|^{2(j_1+j_2+j_3-k)}
|\mu_1|^{2(j_1-j_2-j_3)}
|\mu_2|^{2(j_2-j_1-j_3)}|\mu_3|^{2(j_3-j_1-j_2)} \quad ,
\label{5pt}
\ea
\beq
\mu_1 = \frac{x_{14}\, x_{25}}{z_{14}\, z_{25}} -
\frac{x_{15}\, x_{24}}{z_{15}\, z_{24}} ~, \;\quad
\mu_2 = \frac{x_{15}\, x_{23}}{z_{15}\, z_{23}} -
\frac{x_{13}\, x_{25}}{z_{13}\, z_{25}} ~, \;\quad
\mu_3 = \frac{x_{13}\, x_{24}}{z_{13}\, z_{24}} -
\frac{x_{14}\, x_{23}}{z_{14}\, z_{23}} ~.
\eeq

By performing twice the operation in (\ref{1}) we find the following
expression for the three point function of two $w=1$ and one $w=0$
operators \cite{hmn}
\ba
&&
\!\!\!\!\!\!\!\!\!\!
\left\langle\Phi^{w=1,j_{1}}_{J_{1},{\bar J}_{1}}(x_{1},z_{1})\,
\Phi_{J_{2},{\bar J}_{2}}^{w=1,j_2}(x_{2},z_{2})\,
\Phi_{j_{3}}(x_{3},z_{3})\right\rangle\, \sim\;
W(j_1,j_2,j_3,m_1,m_2,\bar m_1,\bar m_2) \nonumber\\[.8em]
&& \quad\quad \times ~~
\frac{1}{V^{2}_{conf}}\; B(j_1)\, B(j_2)\, C\left(\frac k2-j_1,
\frac k2-j_2, j_3\right) \nonumber\\[.8em]
&& \quad\quad \times ~~
x_{12}^{j_3-J_1-J_2}\;
\bar x_{12}^{\,j_3-\bar J_1-\bar J_2}\;
x_{13}^{J_2-J_1-j_3}\;
\bar x_{13}^{\,\bar J_2-\bar J_1-j_3}\;
x_{23}^{J_1-J_2-j_3}\;
\bar x_{23}^{\,\bar J_1-\bar J_2-j_3}
\nonumber\\[.8em]
&& \quad\quad \times ~~
z_{12}^{\Delta_3-\Delta_1^{w=1}-\Delta_2^{w=1}}\;
\bar z_{12}^{\,\Delta_3-\bar \Delta_1^{w=1}-\bar \Delta_2^{w=1}}\;
z_{13}^{\Delta_2^{w=1}-\Delta_1^{w=1}-\Delta_3}\;
\bar z_{13}^{\,\bar \Delta_2^{w=1}-\bar \Delta_1^{w=1}-\Delta_3}
\nonumber\\[.8em]
&& \quad\quad \times ~~
z_{23}^{\Delta_1^{w=1}
-\Delta_2^{w=1}-\Delta_3}
\bar z_{23}^{\,\bar \Delta_1^{w=1}-\bar \Delta_2^{w=1}-\Delta_3}\; ,
\label{3pww}
\ea
\ba
&&
\!\!\!\!\!\!\!\!\!\!\!\!\!
W(j_1,j_2,j_3,m_1,m_2,\bar m_1,\bar m_2) =
\int d^{2}u\; d^{2}v\; u^{j_{1}-m_{1}-1}\;
{\bar u}^{\,j_{1}-{\bar m}_{1}-1}\;
v^{j_{2}-m_{2}-1}\; {\bar v}^{\,j_{2}-{\bar m}_{2}-1} \nonumber\\[.8em]
&&\qquad\qquad\quad\qquad\qquad\times
|1-u|^{2(j_{2}-j_{1}-j_{3})}\; |1-v|^{2(j_{1}-j_{2}-j_{3})}\;
|u-v|^{2(j_{3}-j_{1}-j_{2})}\; .
\nonumber
\ea
The explicit form of $W$ is known as such integrals were
computed in \cite{fh}. As a successful check, the result in
(\ref{3pww})
was trasformed to the $m$-basis in \cite{4p} using (\ref{max}), thus
giving precisely the
expression for the amplitude of three unflowed operators computed in
\cite{satoh}, up to the powers of $z_{ij}$. This is consistent with the
claim that the coefficient of all winding conserving amplitudes is the
same in the $m$-basis (for a given number of external states)
\cite{malda3,ribault}.

\section{Four Point Functions}

Now we follow \cite{4p} and consider winding conserving four point
functions for operators in
arbitrary spectral flow sectors, both in the $m$- and $x$-basis. We first
recall that the four point
functions of unflowed operators were obtained in \cite{malda3}
by analytically extending to $SL(2,R)$ the results in \cite{tesch2}. We
have
\ba \label{qa}
&&
\!\!\!\!\!\!\!\!\!\!\!\!
\left\langle\Phi_{j_{1}}(x_{1},z_{1})\;
\Phi_{j_{2}}(x_{2},z_{2})\; \Phi_{j_{3}}(x_{3},z_{3})\;
\Phi_{j_{4}}(x_{4},z_{4})\right\rangle \nonumber\\[.8em]
&& = \;\; \int dj\; C(j_{1},j_{2},j)\; B(j)^{-1}\;
C\left(j,j_{3},j_{4}\right)\; {\cal F}(z,x)\;
{\bar{\cal F}}({\bar z},{\bar x}) \\[.8em]
&& \quad\times ~~ |x_{43}|^{2(j_{1}+j_{2}-j_{3}-j_{4})}\;
|x_{42}|^{-4j_{2}}\; |x_{41}|^{2(j_{2}+j_{3}-j_{1}-j_{4})}\;
|x_{31}|^{2(j_{4}-j_{1}-j_{2}-j_{3})} \nonumber\\[.8em]
&& \quad\times ~~
|z_{43}|^{2(\Delta_{1}+\Delta_{2}-\Delta_{3}-\Delta_{4})}\;
|z_{42}|^{-4\Delta_{2}}\;
|z_{41}|^{2(\Delta_{2}+\Delta_{3}-\Delta_{1}-\Delta_{4})}\;
|z_{31}|^{2(\Delta_{4}-\Delta_{1}-\Delta_{2}-\Delta_{3})} ~, \nonumber
\ea
where the integral is over $j=\frac{1}{2}+is$ with $s\in {\bf R}_{>0}$.
Here ${\cal F}$ is a function of the cross ratios
$$x = \frac{x_{21}\,x_{43}}{x_{31}\,x_{42}} ~, \quad\quad
z = \frac{z_{21}\, z_{43}}{z_{31}\, z_{42}} ~,$$
and should be computed by requiring (\ref{qa}) to be a
solution of (\ref{15b}).

Expanding ${\cal F}$ in powers of $z$ as follows
\ba
{\cal
F}(z,x)=z^{\Delta_{j}-\Delta_{1}-\Delta_{2}}\; x^{j-j_{1}-j_{2}}\;
\sum_{n=0}^{\infty}\, f_{n}(x)\, z^{n}\; ,
\label{expan}
\ea
we find $f_{0}$ to obey the hypergeometric equation, so that
we
have the following linearly independent solutions
$${_{2}F_{1}}(j-j_{1}+j_{2},j+j_{3}-j_{4},2j;x)\; ,$$
or
$$x^{1-2j}\; {_{2}F_{1}}(1-j-j_{1}+j_{2},1-j+j_{3}-j_{4},2-2j;\,x)\; .$$
Taking into account both the holomorphic and antiholomorphic parts, the
result in \cite{malda3} was that the unique monodromy invariant
combination is given by
\ba
&&
\!\!\!\!\!\!\!\!\!\!\!\!\!\!\!\!\!\!\!\!\!\!\!\!\!
|{\cal F}(z,x)\;|^2\;=\;
|z|^{2(\Delta_{j}-\Delta_{1}-\Delta_{2})}\;
|x|^{2(j-j_{1}-j_{2})}
{\Bigg\{}\;{\bigg
|}\;{_{2}F_{1}}(j-j_{1}+j_{2},j+j_{3}-j_{4},2j;x)\;{\bigg
|}^{2}\nonumber\\[1em]
&& \qquad +\; \lambda \;
{\bigg
|}\;x^{1-2j}{_{2}F_{1}}(1-j-j_{1}+j_{2},1-j+j_{3}-j_{4},2-2j;x)\;{\bigg
|}^{2}\; {\Bigg \}} +\;\cdots
\label{qaa}
\ea
where the ellipses denote higher orders in $z$ and
$$\lambda = \;-\; \frac{\gamma(2j)^{2}\, \gamma(-j_{1}+j_{2}-j+1)\,
\gamma(j_{3}-j_{4}-j+1)}{(1-2j)^{2}\,\gamma(-j_{1}+j_{2}+j)\,
\gamma(j_{3}-j_{4}+j)}\; .$$
It has been found in \cite{tesch2} that the higher orders in (\ref{expan})
are determined iteratively by the KZ equation starting from $f_{0}$
as the initial condition.

\medskip

Now we would like to extend the results above to the case of winding
conserving four point functions for states in arbitrary spectral flow
sectors, both in the $m$- and $x$-basis. The idea is to first
employ (\ref{max}) in order to transform
(\ref{qa}) to the $m$-basis where we could exploit the fact
that, as mentioned and successfully checked in the previous section, the
coefficient of all winding conserving amplitudes is the same in the
$m$-basis (for a given number of external states). After performing the
spectral flow, we should transform the final result back to the $x$-basis.
In order to simplify the calculations, the only requirement
that we will make is that at least one state is in the spectral flow image
of the highest weight discrete representation. After heavy calculations,
it was found in \cite{4p} that the winding conserving four point functions
for states in arbitrary spectral flow sectors, in the $m$-basis, read
\ba
&&
\!\!\!\!\!
\left\langle\Phi^{w_{1},j_{1}=-m_{1}-n_{1}=-{\bar
m}_{1}-{\bar n}_{1}}_{J_{1},M_{1};{\bar J}_{1}
,{\bar M}_{1}}(z_{1})\;
\Phi^{w_{2},j_{2}}_{J_{2},M_{2};{\bar J}_{2}
,{\bar M}_{2}}(z_{2})\;
\Phi^{w_{3},j_{3}}_{J_{3},M_{3};{\bar J}_{3}
,{\bar M}_{3}}(z_{3})\;
\Phi^{w_{4},j_{4}}_{J_{4},M_{4};{\bar J}_{4}
,{\bar M}_{4}}(z_{4})\right\rangle \nonumber\\[.8em]
&& \quad\sim\;
V_{conf}\; \delta^{2}((m_{1}-n_{1})+m_{2}+m_{3}+m_{4})\;
\frac{\Gamma(2j_{1})^{2}}{\Gamma(j_{1}-m_{1})\,
\Gamma(j_{1}-{\bar m}_{1})} \nonumber\\ &&
\qquad\times ~~
z_{43}^{\Delta^{w_{1}}_{1}(n_{1})+\Delta^{w_{2}}_{2}-\Delta^{w_{3}}_{3}-
\Delta^{w_{4}}_{4}}\;
{\bar z}_{43}^{\,{\bar
\Delta}^{w_{1}}_{1}({\bar n}_{1})+{\bar \Delta}^{w_{2}}_{2}-{\bar
\Delta}^{w_{3}}_{3}-{\bar \Delta}^{w_{4}}_{4}}\;
z_{42}^{-2\Delta^{w_{2}}_{2}}\;
{\bar z}_{42}^{-2{\bar \Delta}^{w_{2}}_{2}}\nonumber\\[.8em]
&& \qquad\times ~~
z_{41}^{\Delta^{w_{2}}_{2}+
\Delta^{w_{3}}_{3}-\Delta^{w_{1}}_{1}(n_{1})-\Delta^{w_{4}}_{4}}\;
{\bar z}_{41}^{\,{\bar \Delta}^{w_{2}}_{2}+
{\bar \Delta}^{w_{3}}_{3}-{\bar
\Delta}^{w_{1}}_{1}({\bar n}_{1})-{\bar \Delta}^{w_{4}}_{4}} \nonumber\\[.8em]
&& \qquad\times ~~
z_{31}^{\Delta^{w_{4}}_{4}-\Delta^{w_{1}}_{1}(n_{1})-\Delta^{w_{2}}_{2}-
\Delta^{w_{3}}_{3}}\;
{\bar z}_{31}^{\,{\bar
\Delta}^{w_{4}}_{4}-{\bar \Delta}^{w_{1}}_{1}({\bar n}_{1})-{\bar
\Delta}^{w_{2}}_{2}-{\bar \Delta}^{w_{3}}_{3}} \nonumber\\[.8em]
&& \qquad\times ~~
\sum_{n_{2},n_{3}=0}^{n_{1}}\;
\sum_{{\bar n}_{2},{\bar n}_{3}=0}^{{\bar n}_{1}}\,
{\cal G}_{n_{2},n_{3}}(j_{i},m_{i})\;
{\cal G}_{{\bar n}_{2},{\bar n}_{3}}(j_{i},{\bar m}_{i})\;
\int dj\; C(j_{1},j_{2},j)\; B(j)^{-1} \nonumber\\[.8em]
&& \qquad\qquad\times ~~ C\left(j,j_{3},j_{4}\right) 
\bigg[\;\Omega (j,j_{i},m_{2}-n_{2},m_{3}-n_{3},
{\bar m}_{2}-{\bar n}_{2},{\bar m}_{3}-{\bar n}_{3}) \nonumber\\[.8em]
&& \qquad\qquad\qquad + \; \lambda\;
\Omega (1-j,j_{i},m_{2}-n_{2},m_{3}-n_{3},
{\bar m}_{2}-{\bar n}_{2},{\bar m}_{3}-{\bar n}_{3})\; \bigg]
\nonumber\\[.8em]
&& \qquad\qquad\times ~~
z^{\Delta^{w}_{j}-\Delta^{w_{1}}_{1}(n_{1})-\Delta^{w_{2}}_{2}}\;
{\bar z}^{\,{\bar \Delta}^{w}_{j}-{\bar
\Delta}^{w_{1}}_{1}({\bar n}_{1})-{\bar
\Delta}^{w_{2}}_{2}}\; + \;\cdots \; ,
\label{corrrr}
\ea
where the ellipses denote higher orders in $z$. We require that this is a
winding conserving correlator, {\it i.e.} we
should have $\sum^{4}_{i=1}w_{i}=0$. The operator at $z_{1}$ is
assumed to be in the spectral flow image
of the highest weight discrete representation, so that we have
introduced the number $n_{1}=0,1,2,\cdots$, which satisfies
$m_{1}=-j_{1}-n_{1}$. We also have
\ba
&&
\!\!\!\!\!\!\!
{\cal G}_{n_{2},n_{3}}(j_{i},m_{i})\equiv\frac{1}{\Gamma(n_{2}+1)
\Gamma(n_{3}+1)}\; \frac{\Gamma(-j_{1}-m_{1}+1)}{\Gamma
(-j_{1}-m_{1}-n_{2}-n_{3}+1)} \nonumber\\[1em]
&&\qquad\quad\quad\qquad\times ~~
\frac{\Gamma
(j_{2}-m_{2}+n_{2})\Gamma
(j_{3}-m_{3}+n_{3})\Gamma(j_{4}-j_{1}-m_{4}-m_{1}-n_{2}-n_{3})}
{\Gamma(j_{2}-m_{2})\Gamma(j_{3}-m_{3})\Gamma(j_{4}-m_{4})}
\; ,\nonumber
\ea
and
\ba
&&
\!\!\!\!\!\!\!\!\!\!
\Omega (j,j_{i},m_{i},{\bar m}_{i}) =\; \Gamma(-j_{3}-j_{4}+j+1)^{2}\;
\Gamma(j_{3}-m_{3})\; \Gamma(j_{3}-{\bar m}_{3}) \nonumber\\[.8em]
&& \!\!\!\times ~~ {\Bigg (}\;
\Lambda\left[ \begin{array}{c}
-j_{1}+j_{2}+j,\; -j_{1}+j_{2}-j+1,\; -j_{1}+j_{2}+j_{4}+m_{3},\; 1\\[.5em]
-j_{1}+j_{2}-j_{3}+j_{4}+1,\; -j_{1}+j_{2}+j_{3}+j_{4},\; j_{2}-m_{2}+1
\end{array} \right] \nonumber\\[.8em]
&& \;+ ~ \Lambda\left[ \begin{array}{c}
-j_{3}-j_{4}+j+1,\; -j_{3}-j_{4}-j+2,\; -j_{3}+m_{3}+1,\;
j_{1}-j_{2}-j_{3}-j_{4}+2\\[.5em]
j_{1}-j_{2}-j_{3}-j_{4}+2,\; -2j_{3}+2,\; j_{1}-j_{3}-j_{4}-m_{2}+2
\end{array} \right] \nonumber\\[.8em]
&& \;+ ~ \Lambda\left[ \begin{array}{c}
j_{3}-j_{4}+j,\; j_{3}-j_{4}-j+1,\; j_{3}+m_{3},\;
j_{1}-j_{2}+j_{3}-j_{4}+1\\[.5em]
j_{1}-j_{2}+j_{3}-j_{4}+1,\; 2j_{3},\; j_{1}+j_{3}-j_{4}-m_{2}+1
\end{array} \right] \; {\Bigg )} \nonumber\\[.8em]
&& \!\!\!\times ~~ {\Bigg (}\;
\Lambda\left [ \begin{array}{c}
-j_{1}+j_{2}+j,\; -j_{1}+j_{2}-j+1,\; -j_{1}+j_{2}+j_{4}+{\bar m}_{3},\; 1\\[.5em]
-j_{1}+j_{2}-j_{3}+j_{4}+1,\; -j_{1}+j_{2}+j_{3}+j_{4},\;
j_{2}-{\bar m}_{2}+1
\end{array} \right] \nonumber\\[.8em]
&& \;+ ~ \Lambda\left [ \begin{array}{c}
-j_{3}-j_{4}+j+1,\; -j_{3}-j_{4}-j+2,\; -j_{3}+{\bar m}_{3}+1,\;
j_{1}-j_{2}-j_{3}-j_{4}+2\\[.5em]
j_{1}-j_{2}-j_{3}-j_{4}+2,\; -2j_{3}+2,\; j_{1}-j_{3}-j_{4}-{\bar m}_{2}+2
\end{array} \right] \nonumber\\[.8em]
&& \;+ ~ \Lambda\left [ \begin{array}{c}
j_{3}-j_{4}+j,\; j_{3}-j_{4}-j+1,\; j_{3}+{\bar m}_{3},\;
j_{1}-j_{2}+j_{3}-j_{4}+1\\[.5em]
j_{1}-j_{2}+j_{3}-j_{4}+1,\; 2j_{3},\;
j_{1}+j_{3}-j_{4}-{\bar m}_{2}+1
\end{array} \right] {\Bigg )} \; ,
\label{oomega}
\ea
with
\ba
&&
\!\!\!\!\!\!\!\!\!\!\!\!\!\!\!\!
\Lambda\left[ \begin{array}{c}
\rho_{1},\rho_{2},\rho_{3},\rho_{4}\\
\sigma_{1},\sigma_{2},\sigma_{3}
\end{array}\right] \equiv\; 
\frac{\Gamma(1-\sigma_{1})\, \Gamma(1-\sigma_{2})\,
\Gamma(1+\rho_{1}-\rho_{2})\, \Gamma(\rho_{4})\,
\Gamma(\sigma_{3} -\rho_{4})}{\Gamma(1-\rho_{2})\,
\Gamma(1-\rho_{3})\, \Gamma(1+\rho_{1}-\sigma_{1})\,
\Gamma(1+\rho_{1}-\sigma_{2})\, \Gamma(\sigma_{3})}
\nonumber\\[1em]
&&\qquad\qquad\qquad\qquad\qquad\qquad
\times ~~ (-1)^{\rho_{1}}\; \displaystyle{_4
F_3(\rho_{1},\rho_{2},\rho_{3},\rho_{4};
\sigma_{1},\sigma_{2},\sigma_{3};1)} \; .
\nonumber
\ea
We complete our analysis by transforming (\ref{corrrr}) back to the
$x$-basis. We take into account that the $x$-basis correlators are the
pole residues of the $m$-basis
results at
$J_{i}=M_{i}$, ${\bar J}_{i}={\bar M}_{i}$, for a given spectral flowed
state. In addition, we point out that the Ward
identities satisfied by
correlators involving either
unflowed or spectral flowed fields in the $x$-basis
are the same up to the replacements
$\Delta_{i}\rightarrow \Delta_{i}^{w},\; j_{i}\rightarrow J_{i}$
for the spectral flowed operators \cite{hmn}, and further, that, to the
lowest order in $z$, the modified
KZ equation (\ref{al2}) actually reduces to that of the
unflowed case with the replacements $j_{i}\rightarrow J_{i}$, and
the iterative terms do not contribute.\footnote{Such a property can
be shown to hold for correlators involving $w=1$ operators, as in
(\ref{al2}). However it seems reasonable to assume that this can be
extended to arbitrary winding number.} Taking all this into account, we
find the following expression in the $x$-basis \cite{4p}
\small
\ba
&&
\!\!\!\!\!\!\!\!\!\!\!\!\!\!\!\!
\left\langle\Phi^{|w_{1}|,j_{1}}_{J_{1}(n_{1}),{\bar J}_{1}({\bar
n}_{1})}(x_{1},z_{1})\;
\Phi^{|w_{2}|,j_{2}}_{J_{2},{\bar J}_{2}}(x_{2},z_{2})\;
\Phi^{|w_{3}|,j_{3}}_{J_{3},{\bar J}_{3}}(x_{3},z_{3})\;
\Phi^{|w_{4}|,j_{4}}_{J_{4},{\bar J}_{4}}(x_{4},z_{4})
\right\rangle \nonumber\\[.8em]
&& \sim ~~ \frac{\Gamma(2j_{1})^{2}}{\Gamma(j_{1}-m_{1})\;
\Gamma(j_{1}-{\bar m}_{1})}\;
x_{43}^{J_{1}(n_{1})+J_{2}-J_{3}-J_{4}}\;
{\bar x}_{43}^{\,{\bar J}_{1}({\bar n}_{1})+{\bar J}_{2}-{\bar J}_{3}-{\bar
J}_{4}}\; x_{42}^{-2J_{2}}\; {\bar x}_{42}^{\,-2{\bar J}_{2}} \nonumber\\[.8em]
&& \quad\times ~~ x_{41}^{J_{2}+J_{3}-J_{1}(n_{1})-J_{4}}\;
{\bar x}_{41}^{\,{\bar J}_{2}+{\bar J}_{3}-{\bar J}_{1}({\bar n}_{1})-{\bar
J}_{4}}\; x_{31}^{J_{4}-J_{1}(n_{1})-J_{2}-J_{3}}\;
{\bar x}_{31}^{\,{\bar J}_{4}-{\bar J}_{1}({\bar n}_{1})-{\bar J}_{2}-{\bar
J}_{3}} \nonumber\\[.8em]
&& \quad\times ~~ z_{43}^{\Delta^{|w_{1}|}_{1}(n_{1})+\Delta^{|w_{2}|}_{2}-
\Delta^{|w_{3}|}_{3}-\Delta^{|w_{4}|}_{4}}\;
{\bar z}_{43}^{\,{\bar \Delta}^{|w_{1}|}_{1}({\bar n}_{1})+{\bar
\Delta}^{|w_{2}|}_{2}-{\bar \Delta}^{|w_{3}|}_{3}-
{\bar \Delta}^{|w_{4}|}_{4}}\; z_{42}^{-2\Delta^{|w_{2}|}_{2}}\;
{\bar z}_{42}^{\,-2{\bar \Delta}^{|w_{2}|}_{2}} \nonumber\\[.8em]
&& \quad\times ~~ z_{41}^{\Delta^{|w_{2}|}_{2}+\Delta^{|w_{3}|}_{3}-
\Delta^{|w_{1}|}_{1}(n_{1})-\Delta^{|w_{4}|}_{4}}\;
{\bar z}_{41}^{\,{\bar \Delta}^{|w_{2}|}_{2}+{\bar \Delta}^{|w_{3}|}_{3}-
{\bar \Delta}^{|w_{1}|}_{1}({\bar n}_{1})-{\bar \Delta}^{|w_{4}|}_{4}}
\nonumber\\[.8em]
&& \quad\times ~~ z_{31}^{\Delta^{|w_{4}|}_{4}-\Delta^{|w_{1}|}_{1}(n_{1})-
\Delta^{|w_{2}|}_{2}-\Delta^{|w_{3}|}_{3}}
{\bar z}_{31}^{\,{\bar \Delta}^{|w_{4}|}_{4}-
{\bar \Delta}^{|w_{1}|}_{1}({\bar n}_{1})-{\bar \Delta}^{|w_{2}|}_{2}-
{\bar \Delta}^{|w_{3}|}_{3}} \nonumber\\[.8em]
&& \quad\times ~~ \sum_{n_{2},n_{3}=0}^{n_{1}}\;
\sum_{{\bar n}_{2},{\bar n}_{3}=0}^{{\bar n}_{1}}\,
{\cal G}_{n_{2},n_{3}}(j_{i},m_{i})\;
{\cal G}_{{\bar n}_{2},{\bar n}_{3}}(j_{i},{\bar m}_{i})\;
\int dj\; C(j_{1},j_{2},j)\; B(j)^{-1} \nonumber\\[.8em]
&& \quad\qquad\times ~~ C\left(j,j_{3},j_{4}\right)\;
\bigg[\;\Omega (j,j_{i},m_{2}-n_{2},m_{3}-n_{3},
{\bar m}_{2}-{\bar n}_{2},{\bar m}_{3}-{\bar n}_{3}) \nonumber\\[.8em]
&& \quad\qquad\qquad +\; \lambda\;
\Omega (1-j,j_{i},m_{2}-n_{2},m_{3}-n_{3},
{\bar m}_{2}-{\bar n}_{2},{\bar m}_{3}-{\bar n}_{3})\; \bigg]\nonumber\\[.8em]
&& \quad\qquad\times ~~ z^{\Delta^{|w|}_{j}-\Delta^{|w_{1}|}_{1}(n_{1})-
\Delta^{|w_{2}|}_{2}}\;
{\bar z}^{\,{\bar \Delta}^{|w|}_{j}-{\bar \Delta}^{|w_{1}|}_{1}({\bar n}_{1})-
{\bar \Delta}^{|w_{2}|}_{2}}\;
x^{j-J_{1}(n_{1})-J_{2}}\; {\bar x}^{j-{\bar J}_{1}({\bar n}_{1})-{\bar
J}_{2}} \nonumber\\[.8em]
&& \quad\qquad\times ~~{\Bigg\{}\; {\hat \lambda}(n_{1}) \;
{\bigg|}\;x^{1-2j}\;\, {_{2}F_{1}}(1-j-J_{1}(n_{1})+J_{2},1-j+J_{3}-
J_{4},2-2j;x)\;{\bigg|}^{2} \nonumber\\[.8em]
&& \quad\qquad\qquad + ~ {\bigg|}\;{_{2}F_{1}}(j-J_{1}(n_{1})+J_{2},j
+J_{3}-J_{4},2j;x)\;{\bigg|}^{2}\; {\Bigg \}} +\;\cdots \; ,
\label{xxbasis}
\ea
\normalsize
where the ellipses denote higher order terms in $z$, and we have
$J_{1}(n_{1})=|-j_1-n_1+\frac k2 w_1|$. Notice that we have replaced
$w_{i}\rightarrow |w_{i}|$ due to the fact that in the $x$-basis
the operators are labeled with positive winding number. We also have

\ba
{\hat \lambda}=\;-\;\frac{\gamma(2j)^{2}\; \gamma(-J_{1}+J_{2}-j+1)\;
\gamma(J_{3}-J_{4}-j+1)}{(1-2j)^{2}\; \gamma(-J_{1}+J_{2}+j)\;
\gamma(J_{3}-J_{4}+j)}\; .
\label{llambda}
\ea

A comment is in order. We emphasize that, whereas the $m$-basis
expression (\ref{corrrr}) holds for all winding conserving four point
functions, including as a particular case the situation in which all the
external operators are unflowed, (\ref{xxbasis}) does not hold when all
the external states are unflowed.
This is consistent with the fact that in the $m$-basis, all $N$ point
functions are the same, for a given $N$,  up to some free boson correlator
which only modifies the $z_i$ dependence.
We point out that the calculations we have performed in order to transform
the amplitude
from the $m$- to the
$x$-basis, involving the evaluation of the pole residue at $J=M$,
${\bar J}={\bar M}$, require at least one
spectral flowed state in the correlator. Thus (\ref{xxbasis}) results in
this case, whereas
(\ref{qa}) holds for four unflowed operators.

We also emphasize that the higher order
contributions to the expansions in $z$
are iteratively determined using modified KZ equations, such as
(\ref{al2}), for
amplitudes involving spectral flowed states,
starting from (\ref{xxbasis}) as the initial condition.

Finally, we point out that one important application of our results
would be to investigate the structure of the factorization of
(\ref{xxbasis}) in order to establish the consistency of
string theory on AdS$_{3}$.

\begin{theacknowledgments}
I would like to thank the organizers for the opportunity to
give this talk, and to E. Herscovich and C. N\'u\~nez for their
collaborations in the papers on which this article is based. This work was
supported by CONICET.
\end{theacknowledgments}

\end{document}